\documentclass[aps,12pt,notitlepage,final,oneside,onecolumn,nobibnotes,
nofootinbib,superscriptaddress,noshowpacs,centertags,amsmath,amssymb]{revtex4}

\begin{document}

\title{Generalized lagrangian of the Rarita--Schwinger field }
\author{\firstname{A.E.} \surname{Kaloshin}}
\email{kaloshin@physdep.isu.ru}
\author{\firstname{V.P.} \surname{Lomov}}
\email{lomov@physdep.isu.ru}
\author{\firstname{A.M.} \surname{Moiseeva}}
% \email{}
\affiliation{Russia, Irkutsk State University}

\begin{abstract}
We derive the most general lagrangian of the free massive
Rarita--Schwinger field, which generalizes the previously known
ones. The special role of the reparameterization transformation is
discussed.
\end{abstract}

\maketitle

\section{Introduction}

The Rarita-Schwinger (R.-S.) vector-spinor field $\Psi^\mu (x)$ is
used for description of the spin-3/2 particles in QFT \cite{Rar}.
But this field has too many degrees of freedom for spin-3/2 and
contains in fact also two spin-1/2 representations.
These "extra" spin-1/2 components are usually supposed to
be unphysical, but the control of this property becomes not so
evident after inclusion of interactions.

In this paper we derive the most general form of the R.-S.
lagrangian without additional assumption about the nature of
spin-1/2 components. Depending on the choice of parameters the
$s=1/2$ components may have the poles in the complex energy plane
or not. The purpose of such construction is twofold. First of all,
it may be useful at renormalization of the dressed R.-S. field
propagator even if we require  the spin-1/2 components to be
unphysical. Another aim may be related with properties of the
hypothetical R.-S. multiplet, corresponding to $\Psi^\mu $ field
and consisting of particles $J^P=3/2^+, 1/2^+, 1/2^-$.

There exist some examples of generalized lagrangians
\cite{Hab,Kir04,Pil} which are constructed by different methods
(usually with using of artificial methods) and for different
purposes. In contrast to previous works our deriving of the
generalized R.-S. lagrangian is a straightforward: we investigate
the propagator instead of equations of motion and utilize
different bases \cite{KL1} to control its properties. Such
approach is more transparent at any step but needs some technical
details, which are collected in Appendix.

\section{Lagrangian}

Free lagrangian of the Rarita-Schwinger field is defined by
differential operator $S^{\mu\nu}$, which is in fact the inverse
propagator
\begin{flalign}
\Lagr=&\overline{\Psi}\vphantom{\Psi}^{\mu}S_{\mu\nu}\Psi^{\nu}.
\label{lagr}
\end{flalign}
 Let us recall the standard choice \cite{Mol} for $S^{\mu\nu}$:
\begin{flalign}
S^{\mu\nu}=&(\hat{p}-M)g^{\mu\nu}+A(\gamma^{\mu}p^{\nu}+
\gamma^{\nu}p^{\mu})+\frac{1}{2}(3A^2+2A+1)\gamma^{\mu}
\hat{p}\gamma^{\nu}+M(3A^2+3A+1)\gamma^{\mu}\gamma^{\nu}.
\label{stand}
\end{flalign}
Here $M$ is the mass of R.-S. field, $p_{\mu}=i
\partial_{\mu}$ and $A$ is an arbitrary real parameter.
Equations of motion, following from \eqref{stand}, lead to
constrains $p\Psi=\gamma\Psi=0$, and therefore to exclusion of the
spin-1/2 degrees of freedom. In other words, the corresponding
terms in propagator should not have poles in energy.

Let us formulate the main requirements for lagrangian:
\begin{enumerate}
\item The fermion lagrangian is linear in derivatives.
\item It should be hermitian $\Lagr^\dag=\Lagr$ \ or \ \
$\gamma^0 (S^{\mu\nu})^\dag\gamma^0=S^{\nu\mu}$.
\item The spin-3/2 contribution has standard pole form
(to be specified below).
\item Lagrangian should not be singular at $p\to 0$. This point is rather
evident but it happens that some rough methods generate
singularities in a propagator, see, e.g. discussion in \cite{BDM}.
\end{enumerate}

 The suitable  starting point to construct the generalized lagrangian
is the most general decomposition of $S^{\mu\nu}$ in
$\gamma$-matrix basis\footnote{We use conventions of Bjorken and
Drell textbook \cite{Bjo}: $\varepsilon_{0123}=1$,
$\gamma^5=\gamma_5=i\gamma^0\gamma^1\gamma^2\gamma^3$ except
that $\sigma^{\mu\nu}=\frac{1}{2}[\gamma^{\mu},\gamma^{\nu}]$.}
\eqref{app:eq1}. The first requirement remains 6 complex
coefficients in (\ref{app:eq1})
\begin{eqnarray}
\label{sec1:eq4} S^{\mu\nu}&=&g^{\mu\nu}\cdot s_{1} +
\hat{p}g^{\mu\nu}\cdot s_{4} + p^{\mu}\gamma^{\nu}\cdot s_{5} +
\gamma^{\mu}p^{\nu}\cdot s_{6} + \sigma^{\mu\nu}\cdot s_{7} + i
\epsilon^{\mu\nu\lambda\rho}\gamma_{\lambda}\gamma^{5}
p_{\rho}\cdot s_{10}=  \\
&=&g^{\mu\nu}(s_{1}-s_{7}) + \hat{p}g^{\mu\nu}(s_{4}-s_{10}) +
p^{\mu}\gamma^{\nu}(s_{5}+s_{10}) + \gamma^{\mu}p^{\nu}
(s_{6}+s_{10}) + \gamma^{\mu} \gamma^{\nu}  s_{7} - \gamma^{\mu}
\hat{p} \gamma^{\nu}  s_{10}. \nonumber
\end{eqnarray}
If we start from the $\gamma$-matrix decomposition with
nonsingular coefficients the fourth requirement is fulfilled
automatically.

 This expression satisfies the condition $\gamma^0
(S^{\mu\nu})^\dag\gamma^0=S^{\nu\mu}$,\ if $s_{1}$, $s_{4}$,
$s_{7}$, $s_{10}$ are real parameters while $s_{5}$ and
$s_{6}=s_{5}^{*}$ may be complex. It is convenient to introduce
the new notations
\begin{equation*}
s_{1}=r_{1},\quad s_{4}=r_{4},\quad s_{7}=r_{7},\quad
s_{10}=r_{10},\quad s_{5}=r_{5}+ i a_{5},\quad s_{6}=r_{5}- i
a_{5},
\end{equation*}
where all parameters are real.

 To take into account the third requirement, we need to
recognize the spin-$3/2$ part of inverse propagator. It is easy to
do in the $\hat{p}$-basis (see Appendix A for details)
\begin{equation}
  \label{sec1:eq5}
S^{\mu\nu}=
(\hat{p}-M)\left(\operP^{3/2}\right)^{\mu\nu}+(\text{spin-$1/2$
contributions}).
\end{equation}
Reversing the Eq.(\ref{sec1:eq5}), we obtain propagator with the
standard pole behavior of spin-3/2 contribution
\begin{equation}
  \label{sec1:eq6}
G^{\mu\nu}= \frac{1}{\hat{p}-M}
\left(\operP^{3/2}\right)^{\mu\nu}+(\text{spin-$1/2$
contributions}).
\end{equation}
Eq.\eqref{sec1:eq5} gives\footnote{We need to use different bases,
  so to distinguish them we use different notations:
  $s_i, S_i, \bar{S}_i$ for coefficients
  in $\gamma$-, $\hat{p}$- and $\Lambda$-basis respectively.}
(using formulae \eqref{app:eq4}, \eqref{app:eq5} for transition
from one basis to another)%
\begin{equation}
S_1=s_1-s_7=-M, \quad S_2=s_4-s_{10}=1 .
\end{equation}
So the $r_{7}, r_{10}$  are dependent values
\begin{equation*}
  \begin{split}
    r_{7}=M + r_{1}, \quad
    r_{10}=r_{4}-1
  \end{split}
\end{equation*}
and we come to four-parameter ($r_{1},r_{4},r_{5},a_{5}$)
lagrangian, which satisfies all the necessary requirements.
\begin{equation}\label{mostl}
S^{\mu\nu}=g^{\mu\nu}(\hat{p}-M) + p^{\mu}\gamma^{\nu}(r_{5}+
r_{4}-1 + i a_{5}) +p^{\nu}\gamma^{\mu}(r_{5}+r_{4}-1 - i a_{5})
+\gamma^{\mu}\gamma^{\nu}(M+r_{1})-
\gamma^{\mu}\hat{p}\gamma^{\nu}(r_{4}-1).
\end{equation}

Other bases may be useful, so let us write down the corresponding
decomposition coefficients (here $E=\sqrt{p^2}$):
\begin{table}[h]
\begin{tabular}{>{$}l<{$}>{$}l<{$}>{$}l<{$}}
   \text{$\gamma$-basis:}     &\quad \text{$\hat{{p}}$-basis:}               &\quad \text{$\Lambda$-basis:}\\
  s_{1}=r_{1},              &\quad S_{1}=-M,                                 &\quad \bar{S}_{1}=-M+E,\\
  s_{2}=s_{3}=0,            &\quad S_{2}=1,                                  &\quad \bar{S}_{2}=-M-E, \\
  s_{4}=r_{4},              &\quad S_{3}=2M+3r_{1},                          &\quad \bar{S}_{3}=(2M+3r_{1})+E(3r_{4}-2),\\
  s_{5}=r_{5}+i a_{5},      &\quad S_{4}=3r_{4}-2,                           &\quad \bar{S}_{4}=(2M+3r_{1})-E(3r_{4}-2),\\
  s_{6}=r_{5}-i a_{5},      &\quad S_{5}=r_{1},                              &\quad \bar{S}_{5}=r_{1}+E(2r_{5}+r_{4}),\\
  s_{7}=M + r_{1},          &\quad S_{6}=r_{4}+2r_{5},                       &\quad \bar{S}_{6}=r_{1}-E(2r_{5}+r_{4}),\\
  s_{8}=s_{9}=0,            &\quad S_{7}=\sqrt{3}\ E(r_{5}-i a_{5}),           &\quad \bar{S}_{7}=\sqrt{3}\ \big[-(M+r_{1})+E(r_{5}-i a_{5})\big],\\
  s_{10}=r_{4}-1,           &\quad S_{8}=-\sqrt{3}\ (M+r_{1})/E,       &\quad \bar{S}_{8}=\sqrt{3}\ \big[(M+r_{1})+E(r_{5}-i a_{5})\big],\\
                            &\quad S_{9}=\sqrt{3}\ E(r_{5}+i a_{5}),           &\quad \bar{S}_{9}=\sqrt{3}\ \big[(M+r_{1})+E(r_{5}+i a_{5})\big],\\
                            &\quad S_{10}=\sqrt{3}\ (M+r_{1})/E,       &\quad \bar{S}_{10}=\sqrt{3}\ \big[-(M+r_{1})+E(r_{5}+i a_{5})\big].
\end{tabular}
\end{table}

\section{Propagator of the Rarita-Schwinger field}

To build the propagator of the Rarita-Schwinger field we need to
reverse \eqref{mostl}
\begin{equation}\label{}
G^{\mu\nu}(p) = \big(S^{-1}\big)^{\mu\nu},
\end{equation}
and the $\Lambda$-basis is convenient here. Reversing of the
spin-tensor leads to set of equations for the scalar coefficients
$\bar{G}_{i}$ \cite{KL1,KL2}.
\begin{gather}
\begin{array}{ccc}
\bar{G}_1 \bar{S}_1 &=& 1, \\
\bar{G}_2 \bar{S}_2 &=& 1,
\end{array}
\notag\\
\begin{array}{ccc}
\bar{G}_3 \bar{S}_3 + \bar{G}_7 \bar{S}_{10}      &=& 1,       \\
\bar{G}_{3} \bar{S}_{7} + \bar{G}_{7} \bar{S}_{6} &=& 0,
\end{array}
\qquad
\begin{array}{ccc}
\bar{G}_4 \bar{S}_4 + \bar{G}_8 \bar{S}_{9}       &=& 1,        \\
\bar{G}_{4} \bar{S}_{8} + \bar{G}_{8} \bar{S}_{5} &=& 0,
\end{array}
\\
\begin{array}{ccc}
\bar{G}_5 \bar{S}_5 + \bar{G}_9 \bar{S}_{8}       &=& 1,        \\
\bar{G}_{5} \bar{S}_{9} + \bar{G}_{9} \bar{S}_{4} &=& 0,
\end{array}
\qquad
\begin{array}{ccc}
\bar{G}_6 \bar{S}_6 + \bar{G}_{10} \bar{S}_{7}      &=& 1,     \\
\bar{G}_{6} \bar{S}_{10} + \bar{G}_{10} \bar{S}_{3} &=& 0 .
\end{array}
\notag
\end{gather}

The equations are easy to solve:
\begin{flalign}
\label{solve}
\bar{G}_1&=\frac{1}{\bar{S}_1},\quad \bar{G}_2=\frac{1}{\bar{S}_2},\notag\\
\bar{G}_3&=\frac{\bar{S}_6}{\Delta_1},\quad
\bar{G}_4=\frac{\bar{S}_5}{\Delta_2},\quad
\bar{G}_5=\frac{\bar{S}_4}{\Delta_2},\quad
\bar{G}_6=\frac{\bar{S}_3}{\Delta_1},\\
\bar{G}_7&=\frac{-\bar{S}_7}{\Delta_1},\quad
\bar{G}_8=\frac{-\bar{S}_8}{\Delta_2},\quad
\bar{G}_9=\frac{-\bar{S}_9}{\Delta_2},\quad
\bar{G}_{10}=\frac{-\bar{S}_{10}}{\Delta_1},\notag
\end{flalign}
where
\begin{equation}
\Delta_1=\bar{S}_{3}\bar{S}_{6}-\bar{S}_{7}\bar{S}_{10},\qquad
\Delta_2=\bar{S}_{4}\bar{S}_{5}-\bar{S}_{8}\bar{S}_{9}.
\label{denom}
\end{equation}
R.-S. propagator in the spin-3/2 sector ($\bar{G}_1, \bar{G}_2$
coefficients) is similar to usual Dirac propagator. As for
spin-1/2 sector ($\bar{G}_3$ -- $\bar{G}_{10}$), it looks like
mixing of two bare propagators with non-diagonal transitions.
Zeros of denominators $\Delta_1, \Delta_2$ are the poles of R.-S.
propagator with quantum numbers $s=1/2$.

Let us write down the denominators following from our lagrangian
\eqref{mostl}:
\begin{align*}
\Delta_{1}(E)&=-M(3M+4r_{1})+
2E(Mr_{5}-Mr_{4}-r_{1})+E^2(-3a_{5}^2-3(r_{4}+r_{5})^2+4r_{5}+2r_{4}),\\
\Delta_{2}(E)&=\Delta_{1}(E\to-E).
\end{align*}
If we want the  spin-1/2 contributions to be unphysical, it
requires $\Delta_{1}=const$  and we come to conditions:
\begin{equation}
  \begin{split}
    M(r_{5}-r_{4})-r_{1}=0&,\\
    3a_{5}^2+3(r_{4}+r_{5})^2-4r_{5}-2r_{4}=0&.
  \end{split}
  \label{nonph}
\end{equation}
One can rewrite it in terms of sum and difference
$\sigma=r_{5}+r_{4}$,\ $\delta=r_{5}-r_{4}$:
\begin{equation}
  \label{nonph'}
  \begin{split}
    r_{1}&=M \delta,\\
    \delta&=3(\sigma^2-\sigma+a_{5}^2) .
  \end{split}
\end{equation}
Let us consider some particular cases of our lagrangian
\eqref{mostl}
\begin{itemize}
  \item To obtain unphysical spin-1/2 sector we should require the
conditions \eqref{nonph}.  If these relations are
fulfilled, we can return to the standard R.-S. lagrangian
\eqref{stand}, if to put $a_{5}=0$ and denote $\sigma=A+1$.
  \item Generalization of the standard lagrangian \eqref{stand}
was suggested by Pilling \cite{Pil}. His lagrangian corresponds to
unphysical $s=1/2$ sector, has two parameters and may be obtained
(in $d=4$ space) from our lagrangian \eqref{mostl} after imposing
of the conditions \eqref{nonph} and changing of notations. Note
that the procedure of deriving \cite{Pil} is used some  trick
\cite{AU} and looks not too transparent.
  \item Lagrangian suggested by Kirchbach and Napsuciale \cite{Kir04}
\begin{equation*}
 S^{\mu\nu}=i\varepsilon^{\mu\nu\lambda\rho}\gamma^5
 \gamma_{\lambda}p_{\rho} - M g^{\mu\nu}
\end{equation*}
corresponds to the choice
\begin{equation*}
  r_4 = r_5 = a_5 = 0, \quad\quad r_1 = -M .
\end{equation*}
It leads to poles in $s=1/2$ sector
\begin{equation*}
  \Delta_1= M(M+2E), \quad \Delta_1= M(M-2E)
\end{equation*}
and presence of these poles contradicts to further analysis
\cite{Kir04} of this lagrangian.
\end{itemize}

\section{Reparametrization}

There exists the well-known transformation of R.-S. field
\begin{equation}
  \label{Field:repar}
  \Psi_{\mu}\to\Psi^{\prime}_{\mu}:\quad \Psi_{\mu}=\theta_{\mu\nu}(B)\Psi^{\prime\nu},
\end{equation}
where $\theta_{\mu\nu}(B)=g_{\mu\nu}+B\gamma_{\mu}\gamma_{\nu}$
and $B=b+i\beta$ is a complex parameter.

This transformation doesn't touch the spin-3/2 because
$(\operP^{3/2})^{\mu\nu}$ operator is orthogonal to
$\gamma^\alpha$. So if to apply it to our inverse propagator
\eqref{mostl}
\begin{equation}\label{trS}
S_{\mu\nu}\to  S^{\ \prime}_{\mu\nu}=\theta_{\mu\alpha}(B^*)
 S^{\alpha\beta} \theta_{\beta\nu}(B) ,
\end{equation}
one can see that $S^{\ \prime}_{\mu\nu}$ keeps all the properties
of $S_{\mu\nu}$ \eqref{mostl}. It means that in fact we have
reparametrization -- after transformation of the \eqref{mostl} we
obtain the same operator with changed parameters
\begin{equation}
  \label{sec1:eq10}
  \theta_{\mu\alpha}(B^*)S^{\alpha\beta}(r_{1},r_{4},r_{5},a_{5})
\theta_{\beta\nu}(B)=S_{\mu\nu}(r_{1}^{'},r_{4}^{'},r_{5}^{'},a_{5}^{'}).
\end{equation}
Direct calculations confirm it, the transformed parameters are the
following:
\begin{equation}
  \begin{split}
    r_{1}^{'}&=r_{1}+2(3M+4r_{1})(2b^2+b+\beta^2),\\
    a_{5}^{'}&=a_{5}(1+4b)+2\beta(2r_{4}+2r_{5}-1),\\
    r_{4}^{'}&=r_{4}+2b^2(4r_{4}-4r_{5}-3)+
    2b(3r_{4}-r_{5}-2)+2\beta^2(4r_{4}-4r_{5}-3)-2\beta a_{5},\\
    r_{5}^{'}&=r_{5}-2b^2(4r_{4}-4r_{5}-3)-2b(r_{4}-
    3r_{5}-1)-2\beta^2(4r_{4}-4r_{5}-3)-2\beta a_{5},
  \end{split}
  \label{repa}
\end{equation}
or, for sum and difference ($\sigma=r_{5}+r_{4}$,\
$\delta=r_{5}-r_{4}$):
\begin{equation*}
  \begin{split}
    \sigma^{'}&=\sigma+2b(2\sigma-1)-4\beta a_{5},\\
    \delta^{'}&=\delta+2(3+4\delta)(2b^2+b+2\beta^2),\\
    a^{'}_{5}&=a_{5}(1+4b)+2\beta(2\sigma-1).
  \end{split}
  \label{repa1}
\end{equation*}

Not so evident but important property of the
$\theta$-transformation is that it doesn't change the pole
positions (masses) of spin-1/2 representations. It may be seen
from the transformation law of denominators $\Delta_1, \Delta_2$
\eqref{denom} of the propagator:
\begin{equation}\label{denomt}
 \Delta_i (E) \to \Delta_i^{\prime}(E)=\Delta_i(E)\cdot \mid 1+4B \mid^2 .
\end{equation}
It means that not all parameters are essential for spectrum of
$s=1/2$ and this fact may be useful for simplification. Applying
the $\theta$-transformation to our $S^{\mu\nu}$ operator
\eqref{sec1:eq10} we can eliminate two of four parameters
\begin{equation*}
  \sigma^{\prime}=0, \quad\quad a_5^{\prime}=0.
\end{equation*}
After some bit of algebra one can find the transformation
parameters
\begin{equation*}
  2b= - \dfrac{\sigma(2\sigma-1)+2a_5^2}{(2\sigma-1)^2+a_5^2},
  \quad\quad 2\beta= \dfrac{a_5}{(2\sigma-1)^2+a_5^2},
\end{equation*}
under the condition
\begin{equation*}
 (2\sigma-1)^2+a_5^2 \neq 0 .
\end{equation*}
Then it is useful to renormalize the R.-S. field\footnote{Recall
that $\theta(a)\theta(b)=\theta(a+b+4ab)$, so inverse
transformation $\theta(a)\theta(\bar{a})=I$ is defined by
parameter $\bar{a}=-a/(1+4a)$.}
\begin{equation}\label{renor}
 \Lagr=\overline{\Psi}\vphantom{\Psi}^{\mu}S_{\mu\nu}\Psi^{\nu}=
 \overline{\Psi}^{\mu}\theta_{\mu\alpha}(\bar{B}^{*})\cdot\theta^{\alpha\beta}(B^{*})
 S_{\beta\rho} \theta^{\rho\lambda}(B)\cdot
 \theta_{\lambda\nu}(\bar{B})\Psi^{\nu}=\overline{\Psi}
 \vphantom{\Psi}^{\prime\alpha}S_{\prime\alpha\lambda}\Psi^{\prime\lambda},
\end{equation}
where $S^{\prime\alpha\beta}$ contains only two parameters
\begin{equation*}
  S^{\prime\alpha\beta}=S^{\alpha\beta}(r_{1}^{'},\delta^{'},\sigma^{'}=0,a_{5}^{'}=0)
\end{equation*}
and renormalized field is
\begin{equation}\label{renor1}
 \Psi^{\prime\lambda}=\theta^{\lambda\nu}(\bar{B})\Psi_{\nu}.
\end{equation}
After field renormalization \eqref{renor1} we have two-parameter
free lagrangian, but now the $\theta$-factor will appear in the
interaction lagrangian through \eqref{Field:repar}. Such
$\theta$-factor with arbitrary parameter (so called "off-shell"
parameter) traditionally exists in interaction lagrangian.
Nevertheless, the meaning of this parameter and its choice is rather
controversial, see, e.g. discussion in \cite{Nat}.

\section{Conclusion}

We obtained the four-parametric lagrangian which satisfies all
general requirements and generalizes all known lagrangians for
Rarita-Schwinger field. We used a straightforward procedure for
its deriving which utilize different bases for spin-tensor
$S^{\mu\nu}(p)$ and studying of propagator instead of equation of
motion. The corresponding propagator has standard form of the
spin-3/2 contribution. As for spin-1/2 terms, they can have poles
in the energy plane which positions and residues depends on the
parameters.

We found that the $\theta$-transformation plays some special role
in our lagrangian: this is a reparametrization which doesn't move
the spin-1/2 poles. We suppose that the meaning of this
degree of freedom may be more clear in case of physical spin-1/2
sector.

We suppose that investigation of the phenomenology of
Rarita-Schwinger multiplet may be interesting and present work is
a necessary step in this direction.

The work was supported in part by RFBR grant  05-02-17722a.

\appendix
\section{Decomposition of spin-tensor}

Propagator or self-energy of the R.-S. field has two spinor and
two vector indices and depends on momentum $p$. We will denote
such object as $S^{\mu\nu}(p)$, omitting spinor indices, and will
call it shortly as a spin-tensor. In our consideration we need to
use different bases for this object.
\begin{enumerate}
\item
Most evident is a $\gamma$-matrix decomposition. It's easy to
write down all possible $\gamma$-matrix structures with two vector
indices. Altogether there are 10 terms in decomposition of
spin-tensor, if parity is conserved.
     \begin{equation}
      \begin{split}
       S^{\mu\nu}(p)=&g^{\mu\nu}\cdot s_1+p^{\mu}p^{\nu}\cdot s_2
       +\hat{p}p^{\mu}p^{\nu}\cdot s_3+\hat{p}g^{\mu\nu}\cdot
       s_4+p^{\mu}\gamma^{\nu}\cdot s_5+
       \gamma^{\mu}p^{\nu}\cdot s_6+ \\
      &+\sigma^{\mu\nu}\cdot s_7+\sigma^{\mu\lambda}p_{\lambda}p^{\nu}\cdot s_8+
       \sigma^{\nu\lambda}p_{\lambda}p^{\mu}\cdot s_9+
       \gamma_{\lambda}\gamma^{5}\imath\varepsilon^{\lambda\mu\nu\rho}p_{\rho}\cdot s_{10}.
      \end{split}
      \label{app:eq1}
      \end{equation}
Here $s_{i}(p^2)$ are the Lorentz-invariant coefficients and
$\sigma^{\mu\nu}=\frac{1}{2}[\gamma^{\mu},\gamma^{\nu}]$.

This is a good starting point of any consideration, since this
basis is complete, nonsingular and free of kinematical constrains.
But this basis is not convenient at multiplication (e.g. in Dyson
summation) because elements of basis are not orthogonal to each
other.
\item
There is another basis (e.g. \cite{Pas95}) for $S^{\mu\nu}$,
which we call as $\hat{p}$-basis. Decomposition of any spin-tensor
in this basis has the form
\begin{eqnarray}
  \label{app:eq2}
S^{\mu\nu}(p)&=&(S_{1}+\hat{p}S_{2})\big(\operP^{3/2}\big)^{\mu\nu}+(S_{3}+
\hat{p}S_{4})\big(\operP^{1/2}_{11}\big)^{\mu\nu}
+(S_{5}+\hat{p}S_{6})\big(\operP^{1/2}_{22}\big)^{\mu\nu}+
\nonumber \\
&&+(S_{7}+\hat{p}S_{8})\big(\operP^{1/2}_{21}\big)^{\mu\nu}+
(S_{9}+\hat{p}S_{10})\big(\operP^{1/2}_{12}\big)^{\mu\nu},
\end{eqnarray}
where  appeared the well-known tensor operator
\cite{Nie,BDM,Pas95}
 \begin{equation}
  \label{app:eq3}
  \begin{split}
    \big(\operP^{3/2}\big)^{\mu\nu}&=g^{\mu\nu}-\big(\operP^{1/2}_{11}\big)^{\mu\nu}-
    \big(\operP^{1/2}_{22}\big)^{\mu\nu},\\
    \big(\operP^{1/2}_{11}\big)^{\mu\nu}&=3\pi^{\mu}\pi^{\nu},\\
    \big(\operP^{1/2}_{22}\big)^{\mu\nu}&=\frac{p^{\mu}p^{\nu}}{p^2},\\
    \big(\operP^{1/2}_{21}\big)^{\mu\nu}&=\sqrt{\frac{3}{p^2}}\cdot\pi^{\mu}p^{\nu},\\
    \big(\operP^{1/2}_{12}\big)^{\mu\nu}&=\sqrt{\frac{3}{p^2}}\cdot p^{\mu}\pi^{\nu},
  \end{split}
 \end{equation}
 which are written here in a non-standard form. Here we introduced
 the "vector"
 \begin{equation}
   \pi^{\mu}=\frac{1}{3p^2}(-p^{\mu}+\gamma^{\mu}\hat{p})\hat{p}
 \end{equation}
 with the following properties:
 \begin{equation}
   (\pi p)=0,\quad (\gamma\pi)=(\pi\gamma)=1,\quad (\pi\pi)=\frac{1}{3},
   \quad \hat{p}\pi^{\mu}=-\pi^{\mu}\hat{p}.
 \end{equation}
\item
The most convenient at multiplication basis is build by combining
the $\operP^{i}_{\mu\nu}$ operators (\ref{app:eq3}) and the
off-shell projection operators
      $\Lambda^{\pm}$
      \begin{equation}
      \Lambda^{\pm}=\frac{1}{2} \left(1 \pm \frac{\hat{p}}{\sqrt{p^2}} \right),
      \label{project}
      \end{equation}
where we assume $\sqrt{p^2}$ to be the first branch of analytical
function. Ten elements of this basis look as
      \begin{flalign}
      \mathcal{P}_{1}=&\Lambda^{+}\mathcal{P}^{3/2},\,&
      \mathcal{P}_{3}=&\Lambda^{+}\mathcal{P}^{1/2}_{11},\,&
      \mathcal{P}_{5}=&\Lambda^{+}\mathcal{P}^{1/2}_{22},\,&
      \mathcal{P}_{7}=&\Lambda^{+}\mathcal{P}^{1/2}_{21},\,&
      \mathcal{P}_{9}=&\Lambda^{+}\mathcal{P}^{1/2}_{12},\notag\\
      \mathcal{P}_{2}=&\Lambda^{-}\mathcal{P}^{3/2},\,&
      \mathcal{P}_{4}=&\Lambda^{-}\mathcal{P}^{1/2}_{11},\,&
      \mathcal{P}_{6}=&\Lambda^{-}\mathcal{P}^{1/2}_{22},\,&
      \mathcal{P}_{8}=&\Lambda^{-}\mathcal{P}^{1/2}_{21},\,&
      \mathcal{P}_{10}=&\Lambda^{-}\mathcal{P}^{1/2}_{12},
      \label{app:eq6}
      \end{flalign}
where tensor indices are omitted. Decomposition in this basis:
\begin{equation}
  \label{app:lambda}
   S^{\mu\nu}(p)= \sum_{A=1}^{10}\bar{S}_{A} \operP_{A}^{\mu\nu}.
\end{equation}

\end{enumerate}

Coefficients of $S^{\mu\nu}$ in $\hat{p}$- and $\gamma$-bases are
related by
\begin{gather}
\begin{array}{ll}
  \label{app:eq4}
    s_{1}=\dfrac{1}{3}(2 S_{1} + S_{3}), \quad\quad &
    s_{2}=\dfrac{1}{3p^2}( -2 S_{1} - S_{3} + 3 S_{5} ),\\
    s_{3}=\dfrac{1}{3p^2}\big( -2S_{2} - S_{4} + 3S_{6} -
    \sqrt{\dfrac{3}{p^2}} (S_{7}+S_{9}) \big),  \quad\quad &
    s_{4}=\dfrac{1}{3}(2 S_{2} + S_{4}),\\
    s_{5}=\dfrac{1}{\sqrt{3p^2}}S_{9},  \quad\quad &
    s_{6}=\dfrac{1}{\sqrt{3p^2}}S_{7},              \\
    s_{7}=\dfrac{1}{3}( - S_{1} + S_{3}),  \quad\quad &
    s_{8}=\dfrac{1}{3p^2}\big( S_{1} - S_{3} -
    \sqrt{3 p^2} S_{8}\big),\\
    s_{9}=\dfrac{1}{3p^2}\big( -S_{1} + S_{3} -
    \sqrt{3 p^2} S_{10}\big),  \quad\quad &
    s_{10}=\dfrac{1}{3}( - S_{2} + S_{4}).
\end{array}
\end{gather}

Reversed relations:
\begin{gather}
\label{app:eq5}
\begin{array}{ll}
    S_{1}=s_{1}-s_{7},                 \quad\quad &
    S_{2}=s_{4}-s_{10},                \\
    S_{3}=s_{1}+2s_{7},                \quad\quad &
    S_{4}=s_{4}+2s_{10},               \\
    S_{5}=s_{1}+p^2s_{2},              \quad\quad &
    S_{6}=p^2s_{3}+s_{4}+s_{5}+s_{6},  \\
    S_{7}=\sqrt{3p^2}s_{6},            \quad\quad &
    S_{8}=-\sqrt{\dfrac{3}{p^2}}( s_{7} + p^2 s_{8}),\\
    S_{9}=\sqrt{3p^2}s_{6},            \quad\quad &
    S_{10}=\sqrt{\dfrac{3}{p^2}}( s_{7}+ p^2 s_{8}).
\end{array}
\end{gather}
Transition from $\hat{p}$- to $\Lambda$-basis:
\begin{flalign}
  \label{app:eq7}
  \bar{S}_{1}&=S_{1}+E S_{2},
  &\bar{S}_{3}&=S_{3}+E S_{4},
  &\bar{S}_{5}&=S_{5}+E S_{6},
  &\bar{S}_{7}&=S_{7}+E S_{8},
  &\bar{S}_{9}&=S_{9}+E S_{10},\notag\\
  \bar{S}_{2}&=S_{1}-E S_{2},
  &\bar{S}_{4}&=S_{3}-E S_{4},
  &\bar{S}_{6}&=S_{5}-E S_{6},
  &\bar{S}_{8}&=S_{7}-E S_{8},
  &\bar{S}_{10}&=S_{9}-E S_{10}.
\end{flalign}
Let us note that $\hat{p}$- and $\Lambda$-bases are singular at
point $p^2=0$. As for branch point $\sqrt{p^2}$ appearing in
different terms, it is canceled in total expression. But poles
$1/p^2$ don't cancel automatically, if we work in $\hat{p}$- or
$\Lambda$-basis. First of all, one can see from \eqref{app:eq5}
that the $S_{7}-S_{7}$ should have kinematical $\sqrt{p^2}$
factors:
\begin{equation}\label{kinfac}
S_{7}=\sqrt{p^2}\tilde{S}_{7},\quad
S_{8}=\frac{1}{\sqrt{p^2}}\tilde{S}_{7},\quad
S_{9}=\sqrt{p^2}\tilde{S}_{9},\quad
S_{10}=\frac{1}{\sqrt{p^2}}\tilde{S}_{10},
\end{equation}
and $\tilde{S}_{i}$ don't have branch point at origin. After it we
see from \eqref{app:eq4} conditions of absence of $1/p^2$ poles:
\begin{equation}
  \begin{split}
    2S_{1}(0)+S_{3}(0)-3S_{5}(0)&=0,\\
    S_{1}(0)-S_{3}(0)-\sqrt{3}\tilde{S}_{8}(0)&=0,\\
    2S_{2}(0)+S_{4}(0)-3S_{6}(0)-\sqrt{3}(\tilde{S}_{7}(0)+\tilde{S}_{9}(0))&=0,\\
\tilde{S}_{8}(0)-\tilde{S}_{10}(0)&=0.\\
  \end{split}
\end{equation}

\section{Equations of motion}

Usually the equations of motion are used for analysis of
Rarita-Schwinger field content. In this approach the "extra"
spin-1/2 components are excluded by additional conditions
$p_{\mu}\Psi^{\mu}=0$, $\gamma_{\mu}\Psi^{\mu}=0$ which should be
the consequences of equations of motion. Let us analyze the
equations of motion, which result from our four-parameter
lagrangian \eqref{mostl}.

 Contracting the  equations of motion with $p_{\mu}$ and
$\gamma_{\mu}$ one can get
\begin{align*}
  p_{\mu}S^{\mu\nu}\Psi_{\nu}&=0,\\
  \gamma_{\mu}S^{\mu\nu}\Psi_{\nu}&=0.
\end{align*}
Our lagrangian \eqref{mostl} leads to the following secondary
constrains:
\begin{equation}
  \label{sec1:eq7}
  \begin{split}
    \Big[(r_{4}+r_{5}-i a_{5})\hat{p}-M\Big]&(p\Psi)+
    \Big[p^2(r_{5}+i a_{5})+(M+r_{1})\hat{p}\Big](\gamma\Psi)=0,\\
    2\Big[2r_{4}+2r_{5}-1+2i a_{5}\Big]&(p\Psi)+\Big[(3M+4r_{1}) +
    (-3r_{4}+r_{5}+2+i a_{5})\hat{p}\Big](\gamma\Psi)=0.
  \end{split}
\end{equation}
This is a system of linear equations (with matrix coefficients) in
$(p\Psi)$, $(\gamma\Psi)$. One can find conditions when this
system has only trivial solution $(p\Psi)=(\gamma\Psi)=0$. Let us
express $(p\Psi)$ from the second equation and substitute to the
first. It gives the equation in $(\gamma\Psi)$:
\begin{equation}
  \label{wave}
  \Delta(p) (\gamma\Psi)=0,
\end{equation}
where $\Delta(p)$ is operator of the form
\begin{multline*}
\Delta(p) = -M (3M+4r_{1}) + \hat{p}\cdot 2 \Big( r_{1}+Mr_{4}-M
r_{5} \Big)+ p^2\Big(-3(r_{4}+r_{5})^2 - 3a_{5}^2 +
2r_{4}+4r_{5}\Big).
\end{multline*}
Eq. \eqref{wave} is in fact some wave equation in the spinor
$(\gamma\Psi)$.  This equation has only trivial solution if
$\Delta(p)=const$, or
\begin{equation}
  \label{sec1:eq9}
  \begin{split}
    r_{1}+Mr_{4}-M r_{5}=0,\\
    3a_{5}^2+3(r_{4}+r_{5})^2-2r_{4}-4r_{5}=0.
  \end{split}
\end{equation}

These relations provide the constrains $(p\Psi)=0$,
$(\gamma\Psi)=0$ and the we can see that the same conditions
\eqref{nonph} guarantee the absence of poles $s=1/2$.


\begin{thebibliography}{100}

\bibitem{Rar}
W.~Rarita and J.~Schwinger. \journal{Phys. Rev.} \textbf{60}
(1941) 61.

\bibitem{Hab}
H.~Haberzettl. nucl-th/9812043.

\bibitem{Kir04}
M.~Kirchbach and M.~Napsuciale. hep-ph/0407179.

\bibitem{Pil}
T. Pilling. \journal{Int.J.Mod.Phys.} \textbf{A20} (2005) 2715.

\bibitem{KL1}
A. E. Kaloshin and V. P. Lomov. \journal{Mod.Phys.Lett.}
\textbf{A19} (2004) 135.

\bibitem{Mol}
P. A. Moldauler and K. M. Case. \journal{Phys. Rev.} \textbf{102}
(1956) 102.

\bibitem{BDM}
M.~Benmerrouche, R.~M.~Davidson and N.~C.~Mukhopadhyay.
\journal{Phys. Rev.} \textbf{C39} (1989) 2339.

\bibitem{Bjo}
J.D.Bjorken and S.D.Drell. Relativistic quantun mechanics.
McGraw-Hill Book Company, 1964.

\bibitem{KL2}
A. E. Kaloshin and V. P. Lomov. hep-ph/0409052

\bibitem{AU}
A.~Aurilia, H.~Umezava. \journal{Phys. Rev.} \textbf{182} (1969)
1686.

\bibitem{Nat}
L.~M.~Nath,B.~Etemadi and J.~D.~Kimel. \journal{Phys. Rev.}
\textbf{D3} (1971) 2153.

\bibitem{Pas95} V. Pascalutsa and O. Scholten.
\journal{Nuc. Phys.} \textbf{A591} (1995) 658.

\bibitem{Nie}
P.~van~Nieuwenhuizen. \journal{Phys. Rep.} \textbf{68} (1981) 189.

\end{thebibliography}
\end{document}